\documentclass[%
aps,prl,showpacs,showkeys,amsmath,amssymb,
twocolumn,
floatfix,
superscriptaddress
]{revtex4-2}

\usepackage{graphicx}
\usepackage{times}
\usepackage{verbatim}
\usepackage{color}
\usepackage{cancel}
\usepackage[normalem]{ulem}
\usepackage{bm}
\usepackage{url}
\usepackage{multirow}
\usepackage[caption=false]{subfig}
\usepackage{ragged2e}
\usepackage{lipsum}

\newcommand{\dmee}{$\Delta m^{2}_{\mathrm{ee}}$}
\newcommand{\thet}{$\sin^{2}2\theta_{13}$}
\newcommand{\nuebar}{$\overline{\nu}_{e}$}

\usepackage{graphicx}
\usepackage{dcolumn}
\usepackage{bm}

\usepackage[dvipsnames]{xcolor}
\usepackage{comment}
\usepackage[version=4]{mhchem}
\definecolor{posurl}{cmyk}{.9 .9 0 0}
\usepackage{hyperref}
\hypersetup{
    colorlinks=true
,urlcolor=posurl
,anchorcolor=posurl
,citecolor=posurl
,filecolor=posurl
,linkcolor=posurl
,menucolor=posurl
,linktocpage=true
,pdfa=true
}

\newcommand{\GEANT}{{\textsc{Geant}}}

\begin{document}

\preprint{APS/123-QED}

\title{Measurement of Electron Antineutrino Oscillation Amplitude and Frequency via Neutron Capture on Hydrogen at Daya Bay}

\newcommand{\IHEP}{\affiliation{Institute~of~High~Energy~Physics, Beijing}}
\newcommand{\Wisconsin}{\affiliation{University~of~Wisconsin, Madison, Wisconsin 53706}}
\newcommand{\Yale}{\affiliation{Wright~Laboratory and Department~of~Physics, Yale~University, New~Haven, Connecticut 06520}} 
\newcommand{\BNL}{\affiliation{Brookhaven~National~Laboratory, Upton, New York 11973}}
\newcommand{\NTU}{\affiliation{Department of Physics, National~Taiwan~University, Taipei}}
\newcommand{\NUU}{\affiliation{National~United~University, Miao-Li}}
\newcommand{\Dubna}{\affiliation{Joint~Institute~for~Nuclear~Research, Dubna, Moscow~Region}}
\newcommand{\CalTech}{\affiliation{California~Institute~of~Technology, Pasadena, California 91125}}
\newcommand{\CUHK}{\affiliation{Chinese~University~of~Hong~Kong, Hong~Kong}}
\newcommand{\NCTU}{\affiliation{Institute~of~Physics, National~Chiao-Tung~University, Hsinchu}}
\newcommand{\NJU}{\affiliation{Nanjing~University, Nanjing}}
\newcommand{\TsingHua}{\affiliation{Department~of~Engineering~Physics, Tsinghua~University, Beijing}}
\newcommand{\SZU}{\affiliation{Shenzhen~University, Shenzhen}}
\newcommand{\NCEPU}{\affiliation{North~China~Electric~Power~University, Beijing}}
\newcommand{\Siena}{\affiliation{Siena~College, Loudonville, New York  12211}}
\newcommand{\IIT}{\affiliation{Department of Physics, Illinois~Institute~of~Technology, Chicago, Illinois  60616}}
\newcommand{\LBNL}{\affiliation{Lawrence~Berkeley~National~Laboratory, Berkeley, California 94720}}
\newcommand{\UIUC}{\affiliation{Department of Physics, University~of~Illinois~at~Urbana-Champaign, Urbana, Illinois 61801}}
\newcommand{\SJTU}{\affiliation{Department of Physics and Astronomy, Shanghai Jiao Tong University, Shanghai Laboratory for Particle Physics and Cosmology, Shanghai}}
\newcommand{\BNU}{\affiliation{Beijing~Normal~University, Beijing}}
\newcommand{\WM}{\affiliation{College~of~William~and~Mary, Williamsburg, Virginia  23187}}
\newcommand{\Princeton}{\affiliation{Joseph Henry Laboratories, Princeton~University, Princeton, New~Jersey 08544}}
\newcommand{\VirginiaTech}{\affiliation{Center for Neutrino Physics, Virginia~Tech, Blacksburg, Virginia  24061}}
\newcommand{\CIAE}{\affiliation{China~Institute~of~Atomic~Energy, Beijing}}
\newcommand{\SDU}{\affiliation{Shandong~University, Jinan}}
\newcommand{\NanKai}{\affiliation{School of Physics, Nankai~University, Tianjin}}
\newcommand{\UC}{\affiliation{Department of Physics, University~of~Cincinnati, Cincinnati, Ohio 45221}}
\newcommand{\DGUT}{\affiliation{Dongguan~University~of~Technology, Dongguan}}
\newcommand{\XJTU}{\affiliation{Department of Nuclear Science and Technology, School of Energy and Power Engineering, Xi'an Jiaotong University, Xi'an}}
\newcommand{\UCB}{\affiliation{Department of Physics, University~of~California, Berkeley, California  94720}}
\newcommand{\HKU}{\affiliation{Department of Physics, The~University~of~Hong~Kong, Pokfulam, Hong~Kong}}
\newcommand{\Charles}{\affiliation{Charles~University, Faculty~of~Mathematics~and~Physics, Prague}} 
\newcommand{\USTC}{\affiliation{University~of~Science~and~Technology~of~China, Hefei}}
\newcommand{\TempleUniversity}{\affiliation{Department~of~Physics, College~of~Science~and~Technology, Temple~University, Philadelphia, Pennsylvania  19122}}
\newcommand{\CGNPG}{\affiliation{China General Nuclear Power Group, Shenzhen}}
\newcommand{\NUDT}{\affiliation{College of Electronic Science and Engineering, National University of Defense Technology, Changsha}} 
\newcommand{\IowaState}{\affiliation{Iowa~State~University, Ames, Iowa  50011}}
\newcommand{\ZSU}{\affiliation{Sun Yat-Sen (Zhongshan) University, Guangzhou}}
\newcommand{\CQU}{\affiliation{Chongqing University, Chongqing}} 
\newcommand{\BCC}{\altaffiliation[Now at ]{Department of Chemistry and Chemical Technology, Bronx Community College, Bronx, New York  10453}} 

\newcommand{\UCI}{\affiliation{Department of Physics and Astronomy, University of California, Irvine, California 92697}} 
\newcommand{\GXU}{\affiliation{Guangxi University, No.100 Daxue East Road, Nanning}} 
\newcommand{\HKUST}{\affiliation{The Hong Kong University of Science and Technology, Clear Water Bay, Hong Kong}} 
\newcommand{\Rochester}{\altaffiliation[Now at ]{Department of Physics and Astronomy, University of Rochester, Rochester, New York 14627}} 

\newcommand{\LSU}{\altaffiliation[Now at ]{Department of Physics and Astronomy, Louisiana State University, Baton Rouge, LA 70803}} 

\newcommand{\NCSF}{\affiliation{New Cornerstone Science Laboratory, Institute of High Energy Physics, Beijing}} 
\author{F.~P.~An}\ZSU
\author{W.~D.~Bai}\ZSU
\author{A.~B.~Balantekin}\Wisconsin
\author{M.~Bishai}\BNL
\author{S.~Blyth}\NTU
\author{G.~F.~Cao}\IHEP
\author{J.~Cao}\IHEP\NCSF
\author{J.~F.~Chang}\IHEP
\author{Y.~Chang}\NUU
\author{H.~S.~Chen}\IHEP
% ...

\collaboration{Daya Bay Collaboration}\noaffiliation

\date{\today}

\begin{abstract}
    This Letter reports the first measurement of the oscillation amplitude and frequency of reactor antineutrinos at Daya Bay via neutron capture on hydrogen using 1958 days of data.
    With over 3.6 million signal candidates, an optimized candidate selection, improved treatment of backgrounds and efficiencies, refined energy calibration, and an energy response model for the capture-on-hydrogen sensitive region, the relative \nuebar~rates and energy spectra variation among the near and far detectors gives $\mathrm{sin}^22\theta_{13} = 0.0759_{-0.0049}^{+0.0050}$ and $\Delta m^2_{32} = (2.72^{+0.14}_{-0.15})\times10^{-3}$ eV$^2$ assuming the normal neutrino mass ordering, and $\Delta m^2_{32} = (-2.83^{+0.15}_{-0.14})\times10^{-3}$ eV$^2$ for the inverted neutrino mass ordering. This estimate of $\sin^2 2\theta_{13}$ is consistent with and essentially independent from the one obtained using the capture-on-gadolinium sample at Daya Bay. The combination of these two results yields $\mathrm{sin}^22\theta_{13}= 0.0833\pm0.0022$, which represents an 8\% relative improvement in precision regarding the Daya Bay full 3158-day capture-on-gadolinium result.
\end{abstract}

\maketitle

The neutrino mixing angle $\theta_{13}$ is one of the independent parameters of the three-neutrino oscillation framework and must be determined experimentally. It is an important input for the determination of other fundamental parameters, most notably the lepton charge-parity violation phase, $\delta_{CP}$~\cite{Workman:2022ynf}, the ordering of the second and third neutrino mass eigenstates~\cite{gonzalez2021nufit}, and the octant of the neutrino mixing angle $\theta_{23}$~\cite{deSalas:2020pgw}. Precise knowledge of this parameter is also critical for flavor model building and for searches of new physics in the neutrino sector, such as non-unitarity of the Pontecorvo–Maki–Nakagawa–Sakata matrix~\cite{Maki:1962mu, pontecorvo1968neutrino,cabibbo1963unitary}.

Precise measurements of $\theta_{13}$ have been reported by the Daya Bay~\cite{DayaBay:2022orm}, RENO~\cite{RENO:2018dro} and Double Chooz~\cite{abe2016measurement} reactor experiments, all of which operate at km-scale baselines and rely on the inverse $\beta$-decay (IBD) reaction, $\overline{\nu}_e+p\to e^++n$, to detect \nuebar's. These experiments were designed to use neutron capture on gadolinium ($n$Gd), $n+\mathrm{Gd} \to \mathrm{Gd}+\gamma s$, as the primary channel to identify IBD neutrons.
The high energy ($\sim$8~MeV) released in $n$Gd capture enhances discrimination from the backgrounds produced by natural radioactivity, the overwhelming majority of which lie below 4~MeV.
Moreover, the detectors use a Gd-loaded liquid scintillator (GdLS) target surrounded by an active liquid scintillator (LS) $\gamma$-catcher volume that significantly mitigates energy leakage which is the undetected energy deposited in non-active volumes.

Oscillation measurements can also be carried out using a sample of IBD events where neutrons capture on hydrogen ($n$H), $n+\ce{^{1}H} \to \ce{^{2}H} + \gamma~(2.2 ~\mathrm{MeV})$. However, the majority of the events in this sample occur in the LS $\gamma$-catcher region, where energy leakage effects are significant. Moreover, there is sizable background from natural radioactivity around the neutron capture signal of $2.2$ MeV leading to larger systematic uncertainties.

These challenges notwithstanding, a precise and virtually independent oscillation measurement can be achieved with the $n$H sample in Daya Bay. Rate-only oscillation measurements of $\theta_{13}$ with the $n$H sample have been reported by this experiment~\cite{an2016new} as well as by RENO~\cite{shin2020observation}. The Double Chooz experiment reported a rate and spectral measurement where all capture targets were simultaneously considered~\cite{de_kerret_double_2020}.

In this Letter, the first measurement of the oscillation amplitude and frequency of reactor \nuebar~at km-scale baselines using only the $n$H channel is reported. The analysis is performed with 1958 days of data collected by Daya Bay from 24 December 2012 to 30 August 2017, resulting in an $n$H sample that is $3.1$ times of the previous one~\cite{an2016new}. Two analyses relying on different calibration approaches, selections, background subtraction techniques, assessment of systematic uncertainties, and fitting methods are employed that yield consistent results. The large statistics, combined with improvements in the understanding of the systematic uncertainties, provides a $\sin^22\theta_{13}$ precision surpassed only by Daya Bay’s $n$Gd result~\cite{DayaBay:2022orm}. An $n$H-only measurement of the mass-squared difference $\Delta m^2_{32}$ is reported for the first time.

The measurement of $\theta_{13}$ presented here has almost no correlation with the one obtained with the $n$Gd sample, as the two \nuebar~samples are distinct and the systematic uncertainties are largely decoupled. The result thus increases confidence in the $n$Gd result and improves the global precision of this important parameter.

In the three-flavor neutrino mixing framework for neutrino propagation in vacuum, which is an excellent approximation to Daya Bay's situation, the probability that a \nuebar~produced with energy $E$ is detected as an \nuebar~after traveling for a distance $L$ is given by
\begin{equation}
    \label{eq:oscillation_prob}
    \begin{aligned}
        P_{\overline{\nu}_{e} \rightarrow \overline{\nu}_{e}} & = 1-\cos ^{4} \theta_{13} \sin ^{2} 2 \theta_{12} \sin ^{2} \Delta_{21} \\                                                 & -\sin ^{2} 2 \theta_{13}\left(\cos ^{2} \theta_{12} \sin ^{2} \Delta_{31}+\sin ^{2} \theta_{12} \sin ^{2} \Delta_{32}\right),
    \end{aligned}
\end{equation}
where $\Delta_{ij} \equiv 1.267 \Delta m_{ij}^{2}~\left(\mathrm{eV}^{2}\right)[L~(\mathrm{m}) / E~(\mathrm{MeV})]$, and $\Delta m_{ij}^2$ is the mass-squared difference between the mass eigenstates $\nu_i$ and $\nu_j$.

Daya Bay uses eight identically-designed Antineutrino Detectors (ADs) to detect \nuebar's~emitted from six $2.9~\mathrm{GW}_\mathrm{th}$ reactor cores.
Four ADs are deployed in pairs in two near experimental halls (EH1 and EH2), where the flux-averaged baseline is about $510~\mathrm{m}$ and $550~\mathrm{m}$, respectively, to constrain the \nuebar~flux at a location where oscillation effects are small. The other four ADs in the far hall (EH3) with a $\sim$1600 m flux-averaged baseline, sample the flux where the term in Eq.~\ref{eq:oscillation_prob} that is modulated by $\sin^2 2\theta_{13}$ is close to maximal.
Each AD consists of three concentric cylindrical tanks: the inner acrylic vessel (IAV) contains 20 tons of $0.1\%$ by weight Gd-doped liquid scintillator (GdLS), the outer acrylic vessel (OAV) 22 tons of undoped liquid scintillator (LS), and the outermost stainless steel vessel (SSV) 36 tons of mineral oil. A total of 192 8-inch diameter photomultiplier tubes (PMTs) are installed in the mineral oil volume on the inner walls of the SSV.
The energy and vertex of each detected event are reconstructed from the charge and time information collected by the AD PMTs. The energy resolution is about $9\%$ at $1$~MeV. The ADs in all halls are surrounded by two optically-decoupled layers of water instrumented with PMTs that identify cosmic-ray muons. More details on the experiment's layout and detectors can be found in Ref.~\cite{an2016detector}.

Measuring the oscillation parameters involves performing a relative comparison of the rate and energy spectrum of reactor \nuebar~at multiple detectors. This requires correcting for instabilities and differences in energy response within each detector (spatial non-uniformities~\footnote{The correction takes into account spatial non-uniformities due primarily to time-dependent changes in the number and location of inoperative channels in individual ADs.}), and establishing a common energy scale for all ADs. The two analyses used in this publication, referred to as A and B, begin from the reconstruction scheme described in Ref.~\cite{an2017measurement}, which relies primarily on spallation neutron (SPN) events tagged by $n$Gd capture to calibrate the energy response and is optimized for the GdLS region.
To improve the performance, most importantly in the LS region, Analysis A calibrates the non-uniform energy response using the $2.2$~MeV signal from $n$H-SPN events.
This analysis uses the fitted 2.2 MeV peak of $n$H-SPN events to align the energy scale of the ADs. The fit is done with a new electromagnetic calorimeter function that decouples energy leakage and detector non-uniformity effects~\cite{Jiahua}.
Analysis B uses the $\alpha$ energy peak from sequential $\beta$-$\alpha$ decays of $^{214}\rm{Bi}$-$^{214}\rm{Po}$ from naturally occurring $^{238}$U and $^{232}$Th to map the non-uniform response. The response is normalized relative to the central GdLS volume so as to rely on the same energy scale calibrations used in Daya Bay's $n$Gd analysis~\cite{DayaBay:2022orm}.
After calibration, the energy scale difference among ADs is less than 0.30\% (0.35\%) for Analysis A (B), and the residual energy response non-uniformity within the ADs is less than $0.5\%$ for each analysis.

The two analyses select IBD events by searching for a prompt positron-like signal in temporal and spatial coincidence with a delayed $n$H-capture-like signal.
Both selections start by removing events caused by spontaneous light emission from the PMTs using a set of improved cuts with negligible IBD detection inefficiency~\cite{DayaBay:2022orm}.
Events close in time to cosmic ray muons are vetoed following the same criteria outlined in Ref.~\cite{an2016new}, and events with reconstructed energies smaller than $1.5$~MeV are likewise rejected to suppress correlated $\beta$-$\alpha$ decays. Surviving prompt and delayed candidates are required to have a reconstructed energy $<12$~MeV and within three times the fitted Gaussian width of each AD's 2.2~MeV neutron capture energy peak, respectively. Prompt-delayed event pairs are required to be within a coincidence time window of [1, 1500] $\mu\mathrm{s}$.
The analyses diverge here, with Analysis A using the same methodology to identify prompt-delayed candidate pairs that has been used in past $n$H analyses described in Refs.~\cite{an2014independent, an2016new, yu2015precise}. Analysis B follows a complementary approach inspired by the $n$Gd IBD selection where prompt and delayed candidates are required to be strictly isolated within the $[1,~1500]~\mu\mathrm{s}$ window~\cite{an2017measurement}.

\begin{figure}[!htb]
    \centering
    \includegraphics[width=8cm]{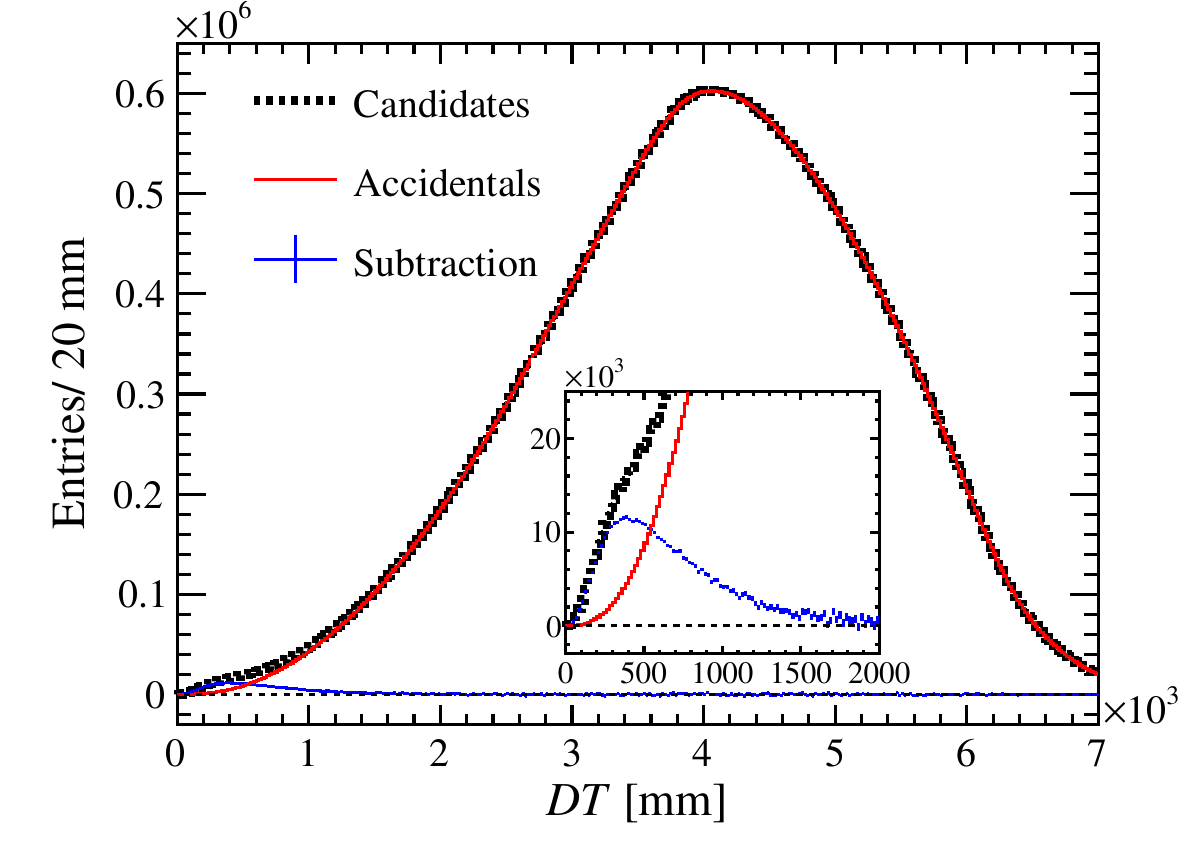}
    \caption{$DT$ distributions of all IBD candidates (black line), synthetic accidentals (red line), and the subtraction of the two (blue points). The latter consists predominantly of true IBD events with a small contamination of correlated backgrounds. A zoom of the $<2000$~mm region is shown in the inset.}
    \label{fig:DT}
\end{figure}

At this point, a large amount of accidental background (explained below) remains given the lower energy released in $n$H capture compared to $n$Gd. Further separation of the signal from the background is achieved by exploiting the tight spatial and temporal correlation of prompt and delayed IBD events. This is the main difference between the $n$H and $n$Gd analyses.
Rather than imposing distinct time and distance separation cuts, as was done previously~\cite{an2014independent, an2016new}, prompt-delayed pairs are required to satisfy a combined $DT\equiv \text{distance}+\text{time}\times v < 1000~\mathrm{mm}$ cut in both analyses, where $v=1000~\rm{mm}/600~\mu\rm{s}$ is close to the speed of thermalized neutrons. This combined cut is more effective at improving the signal-to-background ratio and reduces the relative detection efficiency uncertainty. The $DT$ distribution of EH3 is shown in Fig.~\ref{fig:DT}, which illustrates the importance of this cut as seen from the vast amount of accidental background that would otherwise remain in the IBD candidate sample. The selected $DT$ cut value maximizes the sensitivity to $\theta_{13}$, and the results of both analyses are found to be robust against variations in this cut value.

\begin{table*}[!htb]
    \caption{Summary of signal and backgrounds for Analyses A and B. IBD rates are background-subtracted. All rates are corrected for muon veto and multiplicity selection efficiencies $\varepsilon_{\mu}\times\varepsilon_{m}$. 
    }
    \label{tab:selection_result}
    \begin{minipage}[c]{\textwidth}
        \resizebox{\textwidth}{!}{
            \begin{tabular}{@{\extracolsep{4pt}}lcccccccc@{}}
                \hline\hline
                                                          & EH1-AD1                           & EH1-AD2                           & EH2-AD1                           & EH2-AD2         & EH3-AD1         & EH3-AD2         & EH3-AD3         & EH3-AD4         \\
                \hline
                DAQ live time (days)                      & $1536.624$                        & $1737.620$                        & $1741.214$                        & $1554.046$      & $1739.010$      & $1739.010$      & $1739.010$      & $1551.381$      \\

                \hline
                \multicolumn{9}{c}{Analysis A}                                                                                                                                                                                                                  \\
                IBD candidates                            & $602614$                          & $690506$                          & $688868$                          & $599446$        & $258084$        & $265453$        & $263683$        & $234910$        \\
                $\varepsilon_{\mu}\times \varepsilon_{m}$ & $0.6071$                          & $0.6044$                          & $0.6725$                          & $0.6724$        & $0.9187$        & $0.9179$        & $0.9173$        & $0.9186$        \\
                Accidentals (day$^{-1}$)                  & $119.20\pm0.04$                   & $117.58\pm0.04$                   & $108.47\pm0.03$                   & $104.17\pm0.03$ & $101.28\pm0.03$ & $106.73\pm0.03$ & $105.60\pm0.03$ & $104.78\pm0.03$ \\
                Fast neutron (AD$^{-1}$ day$^{-1}$)       & \multicolumn{2}{c}{$2.78\pm0.33$} & \multicolumn{2}{c}{$2.07\pm0.23$} & \multicolumn{4}{c}{$0.18\pm0.03$}                                                                                           \\
                $^9$Li/$^8$He (AD$^{-1}$ day$^{-1}$)      & \multicolumn{2}{c}{$2.34\pm1.01$} & \multicolumn{2}{c}{$2.83\pm1.15$} & \multicolumn{4}{c}{$0.28\pm0.10$}                                                                                           \\
                Am-C correlated (day$^{-1}$)              & $0.05\pm0.03$                     & $0.05\pm0.03$                     & $0.04\pm0.02$                     & $0.04\pm0.02$   & $0.02\pm0.01$   & $0.02\pm0.01$   & $0.02\pm0.01$   & $0.01\pm0.01$   \\
                Radiogenic neutron (day$^{-1}$)           & \multicolumn{8}{c}{$0.20\pm0.04$}                                                                                                                                                                   \\
                IBD rate (day$^{-1}$)                     & $521.36\pm1.35$                   & $534.49\pm1.33$                   & $474.64\pm1.37$                   & $464.36\pm1.39$ & $59.57\pm0.34$  & $58.88\pm0.34$  & $59.02\pm0.34$  & $59.38\pm0.36$  \\
                \hline
                \multicolumn{9}{c}{Analysis B}                                                                                                                                                                                                                  \\
                IBD candidates                            & $518082$                          & $595250$                          & $619406$                          & $540947$        & $268557$        & $264137$        & $270823$        & $234572$        \\
                $\varepsilon_{\mu}\times \varepsilon_{m}$ & 0.5228                            & 0.5206                            & 0.6028                            & 0.6013          & 0.9183          & 0.9178          & 0.9177          & 0.9180          \\
                Accidentals (day$^{-1}$)                  & $120.05\pm0.06$                   & $119.04\pm0.06$                   & $111.02\pm0.05$                   & $108.12\pm0.06$ & $107.64\pm0.04$ & $106.05\pm0.04$ & $109.74\pm0.04$ & $104.87\pm0.04$ \\
                Fast neutron (AD$^{-1}$ day$^{-1}$)       & \multicolumn{2}{c}{$2.75\pm0.17$} & \multicolumn{2}{c}{$1.97\pm0.10$} & \multicolumn{4}{c}{$0.17\pm0.02$}                                                                                           \\
                $^9$Li/$^8$He (AD$^{-1}$ day$^{-1}$)      & \multicolumn{2}{c}{$3.18\pm1.23$} & \multicolumn{2}{c}{$2.92\pm1.09$} & \multicolumn{4}{c}{$0.19\pm0.07$}                                                                                           \\
                Am-C correlated (day$^{-1}$)              & $0.05\pm0.03$                     & $0.05\pm0.03$                     & $0.04\pm0.02$                     & $0.04\pm0.02$   & $0.02\pm0.01$   & $0.02\pm0.01$   & $0.02\pm0.01$   & $0.01\pm0.01$   \\
                Radiogenic neutron (day$^{-1}$)           & \multicolumn{8}{c}{$0.20\pm0.04$}                                                                                                                                                                   \\
                IBD rate (day$^{-1}$)                     & $518.69\pm1.53$                   & $532.81\pm1.51$                   & $473.96\pm1.33$                   & $465.61\pm1.35$ & $59.89\pm0.34$  & $58.80\pm0.34$  & $59.33\pm0.34$  & $59.21\pm0.35$  \\
                \hline\hline
            \end{tabular}}
    \end{minipage}
\end{table*}

A summary of the IBD candidate, background, and signal rates in each AD can be found in Table~\ref{tab:selection_result}.
It also reports the product of the muon veto and multiplicity efficiencies. The former is precisely determined from the known amount of livetime vetoed, and the latter is estimated with negligible uncertainty from the measured rates of single events~\cite{an2016new,yu2015precise,an2017measurement}.
The dominant background consists of accidental events resulting from the coincidence of two uncorrelated singles predominantly from natural radioactivity. Both analyses generate a synthetic accidental background sample by randomly pairing events from the data that satisfy the IBD selection criteria. The accidental rates are calculated from the measured single rates following the methods of Refs.~\cite{yu2015precise} and \cite{an2017measurement} for Analyses A and B, respectively.
The distance, time, and $DT$ distributions of the synthetic samples match those of selected candidates in the regions where these spatial or temporal separation cuts are significantly relaxed, and the full procedure is also validated using high-statistics simulations~\cite{an2016new, yu2015precise}.
Accordingly, the accidental background subtraction procedure bears negligible systematic uncertainty.

Both analyses perform independent estimations of the fast neutron, Am-C, and $^9$Li/$^8$He backgrounds using the previous methodology~\cite{an2016new}.
The only difference is the estimation of the fast neutron uncertainty. Analysis A compares the rates obtained by using the fitted high energy portion ($>12$ MeV) of the IBD distribution to extrapolate below 12 MeV with a muon-tagged fast-neutron-like sample, whereas Analysis B varies the energy range used for the normalization of the muon-tagged fast-neutron-like sample.

A new correlated background formed by neutrons produced in the spontaneous fission of $^{238}$U or $(\alpha,n)$ reactions in PMT glass was recently identified~\cite{chen2021radiogenic}.
Neutron recoils, or gamma rays produced in fission or de-excitation processes, can mimic the IBD prompt signal. The amounts of natural radioactivity and boron trioxide are measured in several samples of PMT glass with known weight.
A \GEANT4-based simulation is used to estimate the behavior of the neutrons with the fission information obtained from FREYA~\cite{verbeke2018fission} and the ($\alpha,n$) reaction cross-sections obtained from the JENDL/AN-2005 database~\cite{murata2006evaluation}, whose consistency with measurement has been determined to better than $10\%$~\cite{mendoza2020neutron}. The same estimation of this radiogenic neutron background is used for Analyses A and B, as shown in Table~\ref{tab:selection_result}.  Since this background originates in the glass of the PMTs, it is negligible in the GdLS volume and does not affect the $n$Gd analysis~\cite{DayaBay:2022orm} but has a visible impact on this study.

Table~\ref{tab:selection_uncer} summarizes all AD-uncorrelated uncertainties for both analyses.
The total masses of GdLS, LS, IAV, OAV, and mineral oil were measured during the assembly of each AD, and the hydrogen mass fractions were also determined for each chemical component. The uncertainties in these measurements
are propagated to the uncertainty in the efficiency-weighted number of target protons~\cite{an2016new}. The uncertainty introduced by the minimum prompt-delayed time separation cut of $1~\mu\mathrm{s}$ is re-evaluated as in Ref.~\cite{an2016new} and found to remain at $0.10\%$, while the $1500$~$\mu\mathrm{s}$ upper limit has a negligible inefficiency and thus no associated uncertainty.

The application of the 1.5 MeV prompt-energy cut introduces a variation of $0.08\%$ in efficiency between the ADs as a consequence of residual variations in the energy scale after calibration. Variations in energy leakage between the ADs, mainly from geometric differences in non-scintillating volumes (the IAV and the OAV), contribute an additional $0.10\%$, yielding a total uncertainty of $0.13\%$ for the prompt energy cut.

The uncertainty in the fit to the $2.2$~MeV peak, as well as differences among ADs in terms of their $n$H capture ratios, energy resolution, and energy leakage, introduce an AD-uncorrelated uncertainty in the efficiency of the delayed-energy cut that is estimated using the methods first introduced in Ref.~\cite{an2016new}.
Analysis A studies the AD-to-AD differences in ratio of $n$H-SPN to $n$Gd-SPN events, which is particularly sensitive to differences in capture ratio and energy leakage effects, yielding an uncertainty of 0.20\%.
Analysis B compares the ratio of the number of IBD candidates selected within three times the fitted Gaussian width of the 2.2~MeV neutron capture energy peak with the yield of IBD candidates in the larger energy range $[1.5,~2.8]$~MeV to estimate a systematic efficiency difference between ADs of $0.24\%$.
This approach is cross-validated with the $n$H-SPN sample, which has larger statistics and better signal-to-background ratio, yielding consistent results.

For the $DT$ cut, a new IBD sample obtained by relaxing this cut to 3~m and subtracting the synthetic accidental background is used as the denominator to estimate the $DT$ cut efficiency of each AD.
Both analyses estimated the uncertainty by extracting the $1\sigma$ spread of the resulting efficiencies for the eight ADs.
The procedure was repeated on the samples of $^{214}$Bi-$^{214}$Po-$^{210}$Pb cascade decays and $n$H-IBDs with a raised prompt energy cut $E_p>3$ MeV, providing consistent results. The previous estimated uncertainty due to separate distance and time selections~\cite{an2016new} suffered from double-counting of statistical fluctuations leading to a significantly larger systematic uncertainty than that of the current $DT$ cut.

As shown in Table~\ref{tab:selection_uncer}, the total systematic variation in the number of detected \nuebar's between detectors is estimated at $0.34\%$ and $0.37\%$ for Analyses A and B, respectively. Efficiency uncertainties that are correlated among ADs cancel largely due to the relative near-far measurement and are thus not relevant.

\begin{table}[!htb]
    \caption{AD-uncorrelated systematic uncertainties. The row of `Target protons' accounts for relative differences in target mass whereas the next four account for differences in various selection efficiencies among ADs. }\label{tab:selection_uncer}
    \tabcolsep=0.07\linewidth
    \begin{tabular*}{\linewidth}{@{}ccc@{}}
        \hline\hline
        & \multicolumn{2}{c}{Uncertainty ($\%$)}              \\\hline
        & Analysis A                             & Analysis B \\
        Target protons           & $0.11$                                 & $0.11$     \\ 
        Prompt energy            & $0.13$                                 & $0.13$     \\
        $[1,~1500]~\mu$s         & $0.10$                                 & $0.10$     \\
        Delayed energy           & $0.20$                                 & $0.24$     \\
        Coincidence $DT$           & $0.20$                                 & $0.21$     \\
        \hline
        Combined ($\varepsilon$) & $0.34$                                 & $0.37$     \\
        \hline\hline
    \end{tabular*}
\end{table}

\begin{figure}[!htb]
    \centering
    \includegraphics[width=9cm]{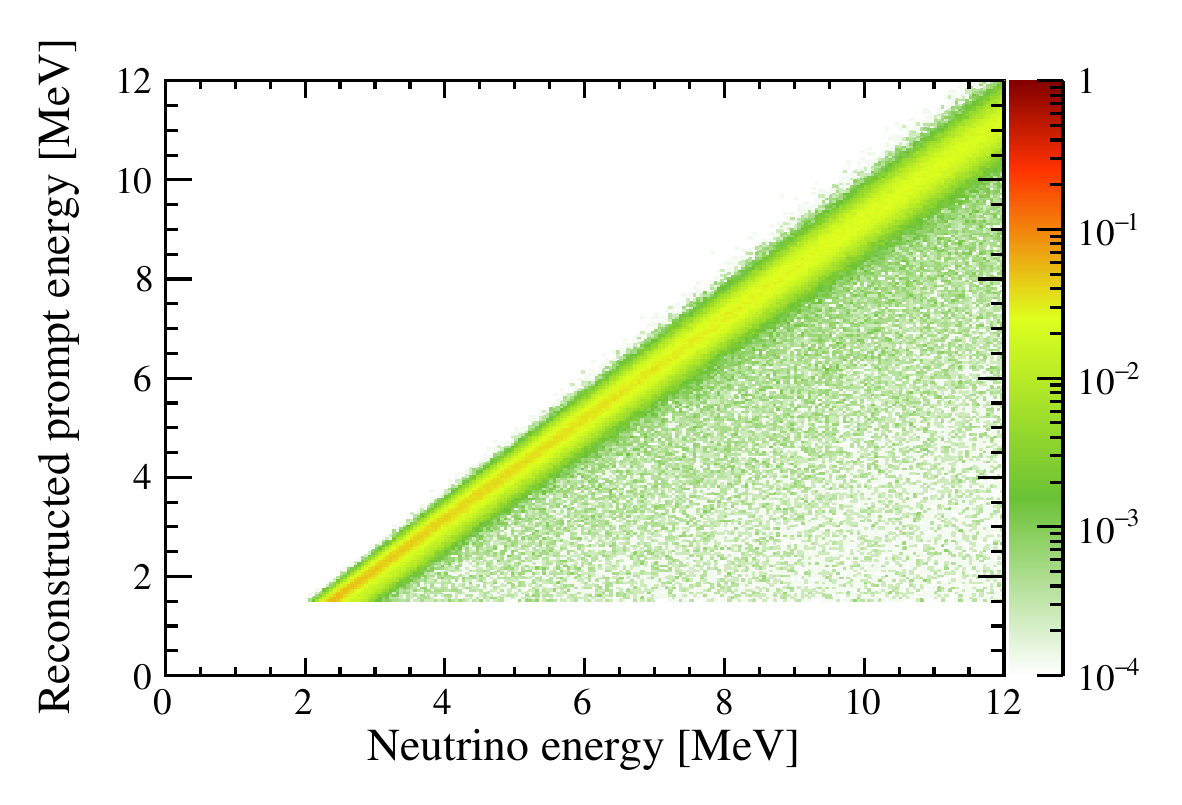}
    \caption{Energy response matrix for an AD, relating the neutrino energy with the reconstructed prompt energy. The matrix is normalized for each neutrino energy bin, according to the actual energy-related selection efficiency. Events in the region away from the diagonal indicates the energy leakage effect.}
    \label{fig:response}
\end{figure}

An energy response model describing the relationship between the true \nuebar~energy and the reconstructed prompt energy, which is shown in Fig.~\ref{fig:response}, is constructed from a full \GEANT4~\cite{agostinelli2003geant4} simulation to be consistent with the data. The model is used by both analyses with some minor differences highlighted below. An extensive calibration campaign including 59 different source and location points is used to tune and benchmark energy leakage effects in the simulation~\cite{DayaBay:2018heb}. The main source of residual uncertainty concerning energy leakage is the slight difference in IAV and OAV geometries between ADs, whose effects are included in the oscillation fits via nuisance parameters. The model also accounts for non-linearities in the conversion of the active energy deposit to scintillation light and the subsequent detection by the PMTs and their associated electronics. These non-linearity effects are measured in the data using multiple $\gamma$ and $\beta$ sources~\cite{adey2019high} and are built directly into the model, with residual uncertainties accounted for via nuisance parameters in the oscillation fits. The cylindrical geometry of the ADs, the non-uniform PMT coverage and the complex PMT angular acceptance induce a non-uniformity in the energy scale and its resolution, particularly in the LS region. The non-uniform energy scale is primarily corrected for in the data using the calibration procedures described earlier. Both analyses corrected the energy scale of the model to match the data, with analysis A applying an additional correction to also match the data in resolution and non-uniformity.
Both analyses use their calibration precision as an uncertainty in their final fits. Finally, the model also accounts for small distortions in the prompt energy spectrum introduced by the $DT$ cut, as a larger cut value, for example, can drive signal vertices to the edge of the AD where the response is slightly different from the center. Analysis B relies entirely on the simulation to correct this effect, while Analysis A uses a data-driven correction derived from a comparison of the energy spectrum obtained with the nominal $DT$ cut versus the one obtained with a loosened 3~m $DT$ cut.

The oscillation parameters are extracted from a $\chi^2$ fit that takes into account the observed \nuebar~rate and spectrum of each AD together with the survival probability of Eq.~\eqref{eq:oscillation_prob}. Analysis A performs a simultaneous fit of the data in all ADs allowing for correlated variations in the rate and spectrum of the reactor \nuebar~prediction to account for any mismodeling. Two independent fitting programs A.1 and A.2 are constructed.
Analysis B follows a strategy similar to Ref.~\cite{DayaBay:2015ayh} where a prediction at the far hall is made directly from the observation at the near halls, minimizing the dependence on reactor \nuebar~prediction models, confirming the results from Analysis A. In both analyses, nuisance parameters are introduced in the $\chi^2$ definition to account for reactor-related, detector-related, and background-related uncertainties.
The best-fit values of both analyses are reported in Table~\ref{tab:bfresults}, where the fitting results of A.1 and A.2 are almost identical and are thus not repeated. The results of Analyses A and B are consistent. There is also good consistency with other estimates, most notably the latest one from Daya Bay using the $n$Gd sample~\cite{DayaBay:2022orm}.

\begin{table}[ht]
    \centering
    \caption{Best-fit results of the two analyses. Rate-only denotes the situation where the fit relies only on the rate differences between the near and far ADs, whereas rate $+$ shape denotes the situation where both the rate and spectral information from all ADs is used. NO (IO) denotes the normal (inverted) neutrino mass ordering. $\Delta m_{\mathrm{ee}}^2$ is the effective mass splitting defined in Ref.~\cite{DayaBay:2019bsz}. }
    \label{tab:bfresults}
    \resizebox{0.49\textwidth}{32mm}{
        \begin{tabular}{ccl}
            \hline\hline
            Analysis           & Strategy                                         & Results                                                                \\
            \hline
            \multirow{8}{*}{A} & \multirow{2}{*}{Rate-only}
                               & $\sin^22\theta_{13}=0.0740^{+0.0068}_{-0.0069}$                                                                           \\
                               &                                                  & $\chi^2/\mathrm{NDF}=4.5/6$                                            \\
                               &                                                                                                                           \\
                               & \multirow{5}{*}{Rate $+$ shape}
                               & $\sin^22\theta_{13}= 0.0759_{-0.0049}^{+0.0050}$                                                                          \\
                               &                                                  & $\Delta m^2_{32}=(2.72^{+0.14}_{-0.15})\times 10^{-3}$ eV$^2$ (NO)     \\
                               &                                                  & $\Delta m^2_{32}=(-2.83^{+0.15}_{-0.14})\times 10^{-3}$ eV$^2$ (IO)    \\
                               &                                                  & $\Delta m_{\mathrm{ee}}^2=(2.77^{+0.14}_{-0.15})\times 10^{-3}$ eV$^2$ \\
                               &                                                  & $\chi^2/\mathrm{NDF}=256.7/236$                                        \\
            \hline
            \multirow{8}{*}{B} & \multirow{2}{*}{Rate-only}
                               & $\sin^22\theta_{13}=0.0714\pm0.0071$                                                                                      \\
                               &                                                  & $\chi^2/\mathrm{NDF}=1.55/3$                                           \\
                               &                                                                                                                           \\
                               & \multirow{5}{*}{Rate $+$ shape}
                               & $\sin^22\theta_{13}= 0.0776\pm0.0053$                                                                                     \\
                               &                                                  & $\Delta m^2_{32}=(2.75\pm0.14)\times 10^{-3}$ eV$^2$ (NO)              \\
                               &                                                  & $\Delta m^2_{32}=(-2.85\pm0.14)\times 10^{-3}$ eV$^2$ (IO)             \\
                               &                                                  & $\Delta m_{\mathrm{ee}}^2=(2.80\pm0.14)\times 10^{-3}$ eV$^2$          \\
                               &                                                  & $\chi^2/\mathrm{NDF}=149.7/134$                                        \\
            \hline\hline
        \end{tabular}}
\end{table}

\begin{figure}[!htb]
    \centering
    \includegraphics[width=9cm]{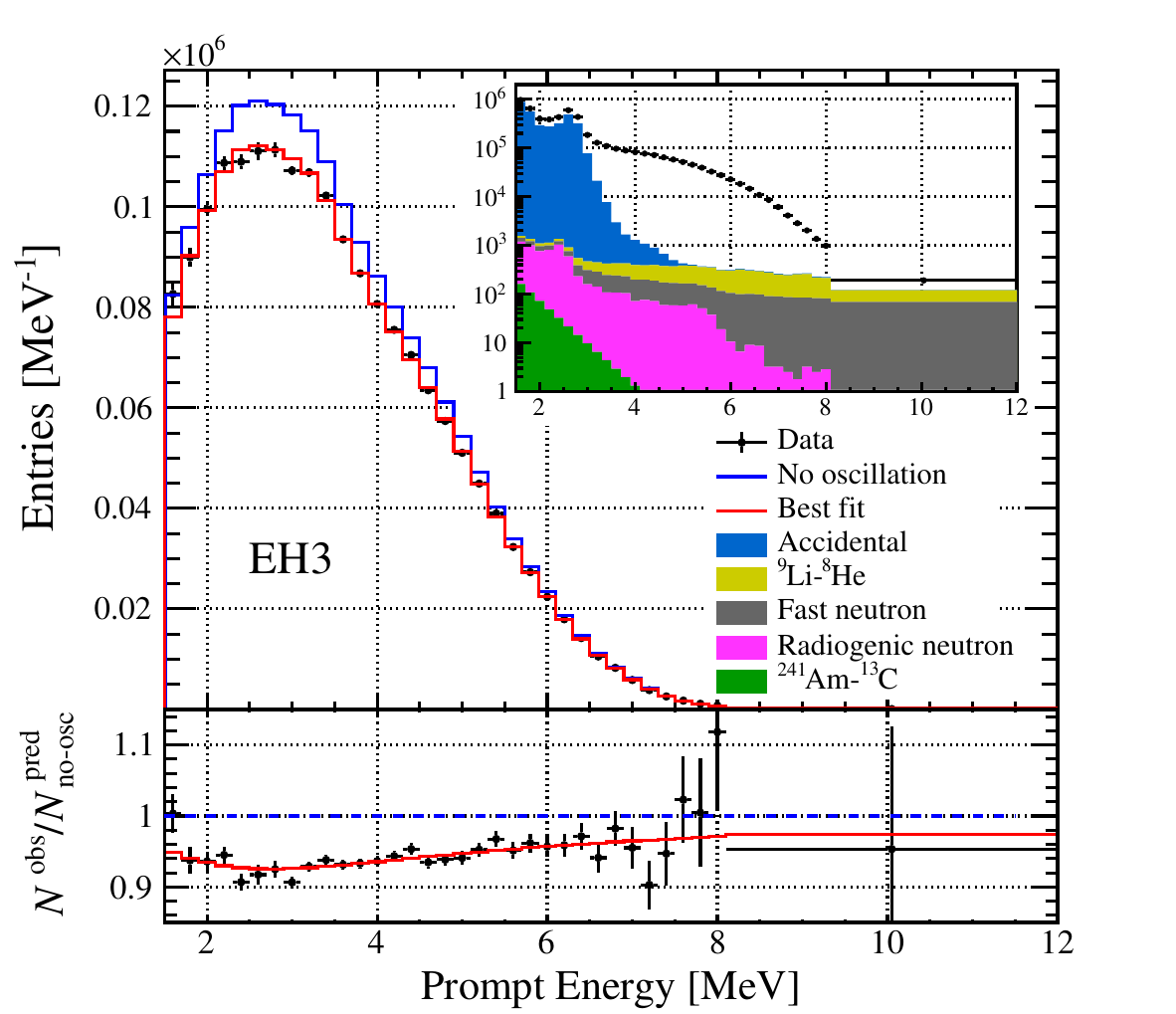}
    \caption{Measured prompt-energy spectrum in EH3 after background subtraction with the best-fit and no-oscillation curves superimposed in the upper panel. The shapes of all candidates and backgrounds are shown in the inset. The lower panel shows the ratio of the observed prompt-spectra after the background subtraction to the predicted no-oscillations one. The error bars are statistical.}
    \label{fig:MvsP_spec}
\end{figure}

\begin{figure}[!htb]
    \includegraphics[scale=.40]{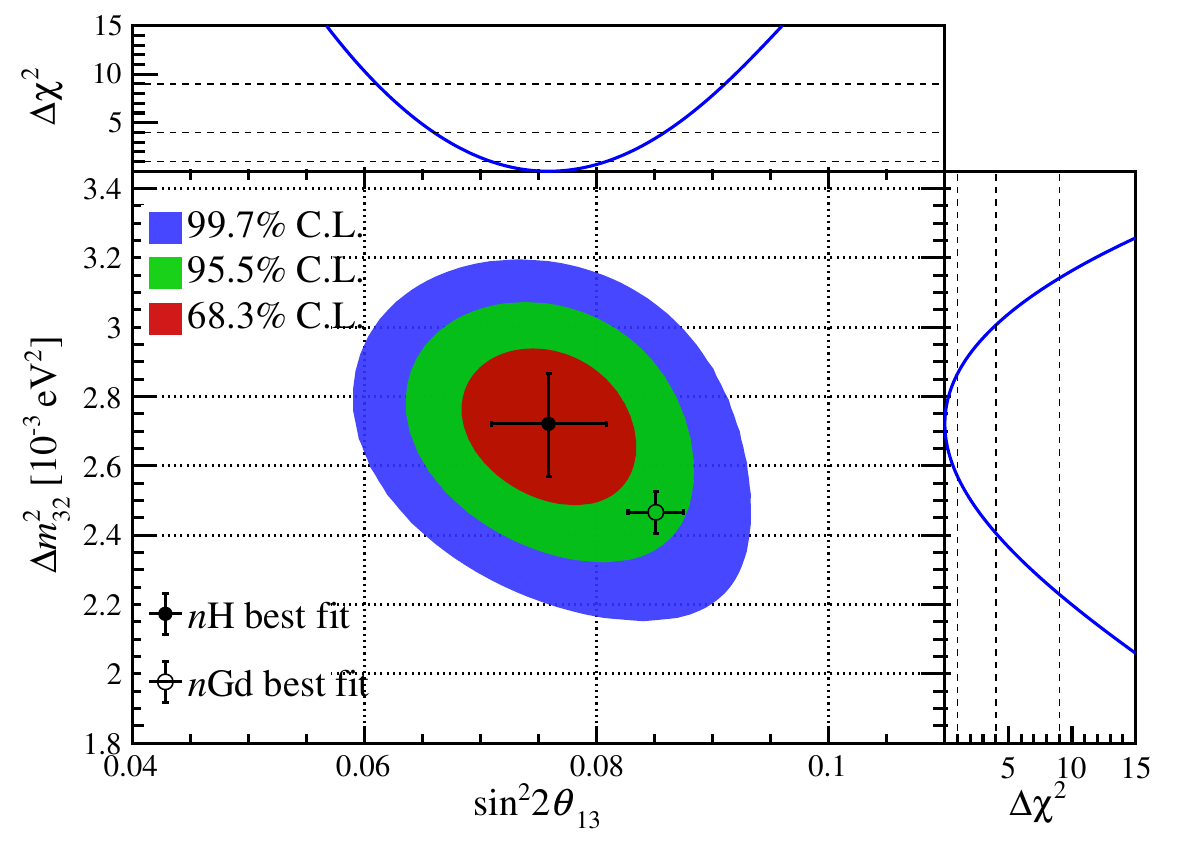}
    \caption{The best estimate of the neutrino oscillation parameters, $\sin^22\theta_{13}$ and $\Delta m_{32}^2$ obtained by exploiting both the rate deficit and shape distortion of the $n$H study are shown together with the allowed regions at different confidence levels. The best estimate from $n$Gd sample~\cite{DayaBay:2022orm} with normal mass ordering is shown too. The error bars include both the statistical and systematic uncertainties.}
    \label{fig:L2EContour}
\end{figure}

The result of Analysis A which achieves a better precision is chosen as nominal.
Figure~\ref{fig:MvsP_spec} shows a comparison between the observed and expected prompt energy spectra in the far hall with no oscillation and with the best-fit oscillation parameters of Analysis A. Fig.~\ref{fig:L2EContour} shows the best-fit result and the $68.3\%$, $95.5\%$, and $99.7\%$ confidence level allowed regions in the $\Delta m^2_{32}$ vs.~$\sin^22\theta_{13}$ plane.
The statistical uncertainty of \thet~and \dmee~accounts for about $47\%$ and $64\%$ of their total uncertainties, respectively.

The significant difference in selection criteria and in signal and background features makes the systematic uncertainties of this $\sin^2 2\theta_{13}$ measurement virtually independent from those of the $n$Gd study.
Using the result from Analysis A, the combined, weighted-average value of $\sin^2 2\theta_{13}$ yields $0.0833\pm0.0022$, which represents an $8\%$ improvement in precision with respect to the $n$Gd estimate alone that was obtained with 3158 days of data~\cite{DayaBay:2022orm}. The combination of $\Delta m^2_{32}$ results suffers from more correlations between the $n$Gd and $n$H analyses, requiring additional study.

In summary, a measurement of reactor \nuebar~disappearance relying on both rate and spectrum is reported using a sample identified via neutron capture on hydrogen collected over 1958 days by the Daya Bay experiment. This work provides an estimate of sin$^22\theta_{13}$ with an uncertainty of about $6.6\%$ that is consistent with and virtually independent of Daya Bay's $n$Gd result~\cite{DayaBay:2022orm}. It also provides the first estimate of $\Delta m^2_{32}$ from Daya Bay using the $n$H sample, which is in good agreement with measurements relying on significantly different neutrino energies, baselines and detection technologies, such as those from accelerator experiments~\cite{abe2021improved,adamson2020precision,acero2021improved,jiang2019atmospheric,aartsen2018measurement}. This result enhances the global precision of both of these parameters and showcases techniques and lessons that can be relevant to other experiments, such as JUNO~\cite{JUNO:2021vlw}, which aims to determine the neutrino mass ordering and other oscillation parameters to high precision using the $n$H channel.

Daya Bay is supported in part by the Ministry of Science and
Technology of China, the U.S. Department of Energy, the Chinese
Academy of Sciences, the CAS Center for Excellence in Particle
Physics, the National Natural Science Foundation of China, the
Guangdong provincial government, the Shenzhen municipal government,
the China General Nuclear Power Group, Key Laboratory of Particle and
Radiation Imaging (Tsinghua University), the Ministry of Education,
Key Laboratory of Particle Physics and Particle Irradiation (Shandong
University), the Ministry of Education, Shanghai Laboratory for
Particle Physics and Cosmology, the Research Grants Council of the
Hong Kong Special Administrative Region of China, the University
Development Fund of The University of Hong Kong, the MOE program for
Research of Excellence at National Taiwan University, National
Chiao-Tung University, and NSC fund support from Taiwan, the
U.S. National Science Foundation, the Alfred~P.~Sloan Foundation, the
Ministry of Education, Youth, and Sports of the Czech Republic,
the Charles University Research Centre UNCE,
the Joint Institute of Nuclear Research in Dubna, Russia, the CNFC-RFBR
joint research program, the National Commission of Scientific and
Technological Research of Chile, and the Tsinghua University
Initiative Scientific Research Program. We acknowledge Yellow River
Engineering Consulting Co., Ltd., and China Railway 15th Bureau Group
Co., Ltd., for building the underground laboratory. We are grateful
for the ongoing cooperation from the China General Nuclear Power Group
and China Light and Power Company.

\bibliographystyle{apsrev4-1}
\bibliography{ref}

\end{document}